\documentclass[aps,prx,reprint,superscriptaddress,longbibliography]{revtex4-2}
\usepackage[normalem]{ulem}
\usepackage{amsmath}
\usepackage{amssymb}
\usepackage{graphicx}
\usepackage{xcolor}
\usepackage{subcaption}
\usepackage{comment}
\usepackage{orcidlink} 
\usepackage{CJKutf8}   

\definecolor{pyang}{rgb}{0.83, 0.69, 0.22}
\newcommand{\bs}{\boldsymbol}
\graphicspath{{./figures/}}

\begin{document}
\title{Tunable Topological Charge Order in Magnetic Moir\'e Systems}
\author{Conor Smith \orcidlink{0009-0005-4580-762X}}
\affiliation{Center for Computational Quantum Physics, Flatiron Institute, New York, NY, 10010, USA}
\author{Yubo Yang (\begin{CJK*}{UTF8}{gbsn}杨煜波\end{CJK*}) \orcidlink{0000-0002-8800-9426}}
\affiliation{Center for Computational Quantum Physics, Flatiron Institute, New York, NY, 10010, USA}
\affiliation{Department of Physics and Astronomy, Hofstra University, Hempstead, NY, 11549, USA}
\author{Shiwei Zhang \orcidlink{0000-0001-9635-170X}}
\affiliation{Center for Computational Quantum Physics, Flatiron Institute, New York, NY, 10010, USA}
\author{Miguel A. Morales \orcidlink{0000-0002-6389-3067}}
\affiliation{Center for Computational Quantum Physics, Flatiron Institute, New York, NY, 10010, USA}

\date{\today}

\begin{abstract}
Two-dimensional electrons in a strong magnetic field exhibit two seemingly
incompatible ground states: incompressible quantum Hall liquids with a
quantized Hall conductance, and electron crystals in which interactions
drive the charge into ordered patterns. Moir\'e superlattices in
transition-metal-dichalcogenide bilayers offer a tunable arena for this
competition. Using neural quantum states (NQS) together with several methodological advances, we show how this competition unfolds. 
In particular, we 
demonstrate that the two end points are separated by an extended regime in which molecular charge order coexists with a quantized Hall response. We
map the ground-state phase diagram
of electrons at Landau-level filling $\nu_B=1$ in the presence of a moir\'e
superlattice potential, with three electrons
per moir\'e cell. 
In the deep triangular potential limit, the electrons form the topologically trivial
``Wigner molecular crystal'' of three-electron artificial atoms, a state recently
imaged at zero magnetic field in moir\'e
bilayers. As the potential depth is lowered, molecular charge order remains while quantized Hall
response emerges. At the
honeycomb point, where each cell contains three electrons and two degenerate moir\'e minima, wells are either singly or doubly occupied and the doublons provide an emergent Ising degree of freedom. Starting from modest interaction strength, we find short-range doublon correlations within the Chern insulator which grow with interaction strength and eventually break sublattice symmetry, forming a N\'eel CDW with long-range order between doublons. The topological molecular regime lies at twist angles and magnetic fields directly accessible to current experiments. The methodological advances presented provide a scalable framework for exploring a broad range of questions in correlated quantum matter in the presence of magnetic fields.

\end{abstract}

\maketitle

\section{Introduction}
Strong magnetic fields and strong Coulomb repulsion provide different organizational principles for a two-dimensional (2D) electron system. Landau quantization quenches the kinetic energy and favors incompressible quantum Hall liquids with quantized Hall response~\cite{Klitzing1980,TsuiStormer1982,Laughlin1983}, while interactions favor crystallization into electron solids~\cite{Wigner1934,Fukuyama1979,LamGirvin1984,Levesque1984,Sung_ElectronicMicroemulsionPhase_2025}, realized as insulating Wigner crystals at low Landau-level filling~\cite{Andrei1988,Tsui2024} and as bubble and stripe phases in higher Landau levels~\cite{Koulakov1996,MoessnerChalker1996,Lilly1999,Cooper1999}.
Although they compete, these two organizing principles need not be mutually exclusive. Te\v{s}anovi\'{c}, Axel, and Halperin proposed the ``Hall crystal''~\cite{Tesanovic1989}, an electron solid whose occupied band carries a nonzero Chern number. 
Topological electron crystals have since been reported in moir\'e materials at zero magnetic field, where charge order coexists with a quantized Hall response~\cite{Polshyn2022,Cai2023,Park2023,XuPRX2023,Zeng2023,Kang2024,Lu2024,DongPatri2024,DongAHC2024,Soejima2024,valenti2025quantumgeometrydrivencrystallization,Lu2025,Su2025}. In all of these zero-field systems, however, the Berry curvature responsible for the Hall response is inherited from the host band structure, so charge order, band topology, and lattice geometry are intertwined and difficult to tune independently.

Transition-metal-dichalcogenide (TMD) moir\'e materials make the organizational principle of crystallization directly visible. A twist or lattice mismatch between layers produces a long-wavelength periodic potential in which interactions can stabilize generalized Wigner crystals~\cite{Regan2020,XuNature2020,Li2021}, whose melting and response to disorder have been imaged directly~\cite{Xiang2025,ge2025visualizingimpactquencheddisorder}. For example, at a filling of three electrons per moir\'e cell, the electrons in each well bind into a three-electron ``artificial atom'', forming Wigner molecular crystals that have been realized experimentally~\cite{li2024wigner} and modeled with neural quantum states~\cite{Li2025,khalifa2026continuumneuralquantumstates}.

An applied magnetic field separates these ingredients. Landau levels are exactly flat and each carries unit Chern number so topology is supplied by the field directly. The moir\'e potential independently supplies the localization and the superlattice reorganizes the levels into Hofstadter minibands~\cite{Hofstadter1976,Dean2013} whose topology is constrained by the flux threading each cell. 
Interactions within these bands then stabilize Chern insulators. Landau-level mixing can by itself drive crystallization~\cite{JainDMC}, and a moir\'e potential can pin electrons into solids that need not be topologically trivial~\cite{Reddy_Qmelt-WC-Hall_2025}. Placing highly correlated states like the Wigner molecular crystals in such a field poses a question zero-field platforms cannot cleanly isolate: when Landau quantization and moir\'e scale localization are present in a single system, under what conditions does the molecular charge order induced by the moir\'e potential destroy or preserve the quantized Hall response due to Landau quantization?

In this work, we address this question for interacting electrons in the continuum at Landau-level filling $\nu_B=1$ and moir\'e filling $\nu_m=3$, corresponding to three magnetic flux quanta per moir\'e unit cell. Using parameters appropriate to electron-doped TMD heterobilayers, we tune both the depth of the moir\'e potential, $V_m$, and its shape, parameterized by $\phi$, which interpolates continuously between a honeycomb potential with two symmetry-related minima per cell ($\phi=60^\circ$) and a triangular potential with a single minimum ($\phi=0^\circ$)~\cite{Wu2018,Yang_MetalInsulatorTransitionSemiconductor_2024,Yang_FerromagneticSemimetalChargeDensity_2024}. At the densities of interest, the Coulomb energy at the magnetic length exceeds the cyclotron energy by an order of magnitude, so Landau-level mixing is strong and the problem lies beyond the reach of lowest-Landau-level-projected methods.

We find that the integer quantum Hall liquid does not give way directly to a trivial electron solid. Instead, the many-body Chern number remains $C=1$ over an extended intermediate regime, where the charge density becomes strongly organized within the moir\'e unit cell. The pattern of charge in this topological regime is controlled by the moir\'e geometry: near the honeycomb limit both sublattices remain occupied in a (2:1) arrangement, while increasing triangular bias collapses the charge onto a single sublattice in a (3:0) arrangement reminiscent of the Wigner molecular crystals observed at zero field~\cite{li2024wigner}. The eventual loss of the quantized response is not only tied to the degree of localization but also to the moir\'e geometry. Near the triangular limit, the trivial states emerge in fairly shallow moir\'e potentials, while in the honeycomb limit
the quantized Hall response survives to the largest potential depths we study. More strikingly, the honeycomb state is unique within the phase diagram. With three electrons distributed over two equivalent minima per moir\'e unit cell, each well is either singly or doubly occupied. Further, since the electrons are spin-polarized, the double occupancy is not Hubbard-like but a correlated two-electron doublon. These doublons provide an emergent Ising degree of freedom at this filling. They repel one another and alternate locally with singly occupied sites. At moderate interaction strength, this alternation is short-range and the state retains a Chern number of $C=1$, which we refer to as a Chern insulator with doublon correlations.  We investigate the fate of the doublon ordering versus interaction strength and find that at strong interaction, the state breaks sublattice symmetry and forms a N\'eel charge-density wave (CDW) of doublons.

\begin{figure*}[ht]
\centering
  \includegraphics[width=\linewidth]{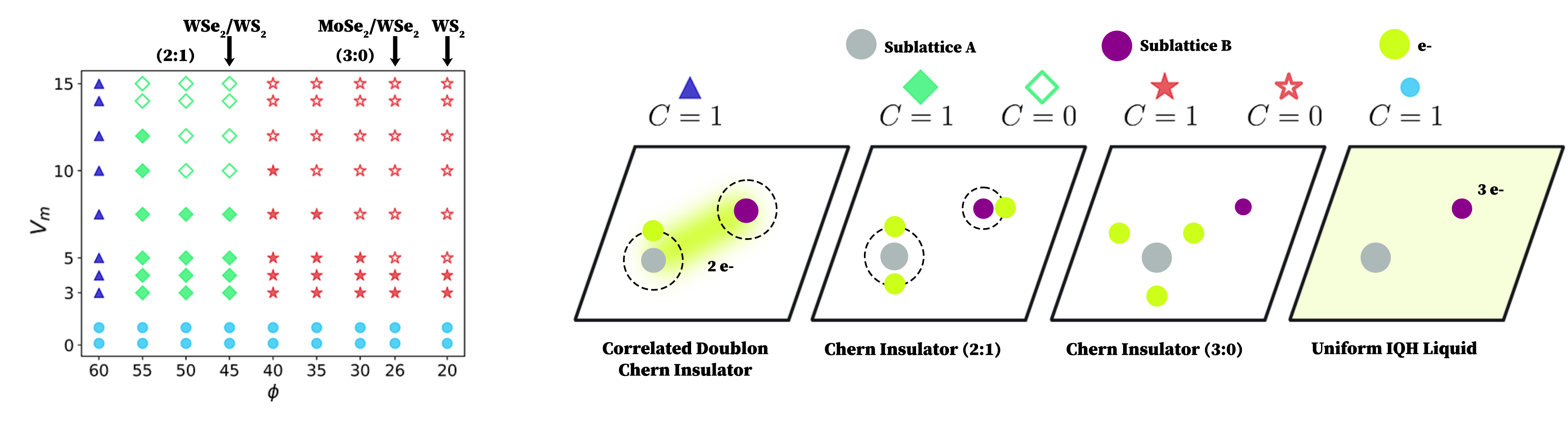}
\caption{
Summary of main physics results: schematic ground-state phase diagram for the moir\'e system in a magnetic field at magnetic filling $\nu_B=1$, moir\'e filling $\nu_m=3$ and interaction strength $r_s=15$, as a function of lattice geometry and moir\'e potential depth. Three systems already realized experimentally are indicated by arrows. Panels on the right illustrate the different states represented by the different symbols in the phase diagram, distinguished by their many-body Chern number, sublattice occupation and density profiles.
The grey and purple circles represent the locations of the moir\'e potential minima, with their relative size indicating the depths of the potential of the two sublattices. Green circles indicate the location (and spread) of the electrons.
}
\label{fig:1}
\end{figure*}

Our results are summarized in Fig.~\ref{fig:1}.
The parameter regime we study was chosen to be within current
experimental reach. The moir\'e shapes highlighted correspond to real
materials---twisted bilayer WS$_2$ realizes a strongly triangular potential
($\phi\approx20^\circ$)~\cite{li2024wigner}, MoSe$_2$/WSe$_2$ a weakly
triangular one ($\phi\approx26^\circ$)~\cite{Wu2018}, and WSe$_2$/WS$_2$
an intermediate shape
($\phi\approx45^\circ$)~\cite{Regan2020,Li2021,Yang_moire-quantum-chemistry_2020}---and the
required fields, $B\approx12$--$47$~T at twist angles of
$0.5^\circ$--$1.1^\circ$, overlap with DC magnet capabilities for the
larger moir\'e periods. The topological analog of the molecular Wigner crystal we predict (the (2:1) and (3:0) ``Chern insulators'' with $C=1$) would
appear as a quantized Hall plateau coexisting with the molecular charge
patterns already imaged in zero field experimental devices~\cite{li2024wigner}, while the honeycomb
state offers a complementary target. Its doublon correlations are invisible in the average density and appear only in density-density correlations. Even the N\'eel CDW of doublons at high interaction strength restores symmetry over both sublattices in the density channel. However, a slight strain in the sample or sublattice bias would encourage the state to break symmetry in the density and allow for scanning-probe imaging. This work thus provides a direct roadmap for existing zero-field imaging experiments to explore the magnetic-field regime.

Mapping this phase diagram requires solving for the ground state of a strongly correlated continuum problem with full Landau-level mixing, together with a direct probe of many-body topology. We meet both requirements with 
neural quantum states (NQS)~\cite{Carleo2017}, which have recently reached the accuracy frontier for continuum electron gases~\cite{Pescia_2024,Smith_2024,smith2026neuralnetworkdiscoverypaired} and have begun to address quantum Hall physics~\cite{Teng2025,Qian2025,sb8g-qvn6,9qlr-jp6x}. Our computational approach is illustrated in Fig.~\ref{fig:NN}. As we detail below in Sec.~\ref{sec:method}, this is made possible by three methodological elements: (i) an ansatz in which a fully periodic multiplicative orbital backflow acts on a Landau-level basis, encoding correlations directly at the many-body level while satisfying magnetic periodic boundary conditions by construction; 
(ii) direct measurement of the many-body Chern number via flux insertion~\cite{Laughlin1981,Niu1985}, evaluated as the winding of the Resta polarization~\cite{Resta_1998} under twisted boundary conditions; and (iii) a foundation model~\cite{Rende_2025} in which the Hamiltonian parameters $(V_m,\phi)$ and the boundary twist enter as inputs to the network, so that a single trained model represents ground states across the entire phase diagram and renders affordable the many ground-state evaluations required for topological characterization. Beyond the specific phase diagram established here, this combination turns NQS into a practical instrument for discovering and diagnosing topological order in continuum quantum matter.

The remainder of this paper is organized as follows. Section~\ref{sec:model} defines the heterobilayer moir\'e continuum Hamiltonian in a magnetic field and its mapping to TMD material parameters. Section~\ref{sec:method} presents the methodological components: the neural network multiplicative orbital backflow (NN-MOB) ansatz on a Landau-level basis (Sec.~\ref{sec:prefactor}), parameter conditioning and the direct measurement of many-body Chern numbers (Sec.~\ref{sec:chern}), and the local observables used to characterize charge order. Section~\ref{sec:results} presents the phase diagram, the interplay of charge order and topology, the doublon correlations and their ordering transition at the honeycomb point, and the topological-to-trivial transitions. Sections~\ref{sec:discussion} and~\ref{sec:conclusion} discuss the implications and experimental signatures.

\section{Model}
\label{sec:model}
Using the Wigner-Seitz radius $a$ as length unit and the kinetic energy scale $W=\frac{\hbar^2}{m^*a^2}$ as energy unit
($m^*$ is the effective mass of the electron),
the interacting moir\'e continuum Hamiltonian in a perpendicular magnetic field can be written as
\begin{equation} \label{eq:mch-rs}
\mathcal{H} = \frac{1}{2}\sum\limits_i \Pi_i^2 - \lambda\sum\limits_i \Lambda(\bs{r}_i) + r_s\sum\limits_{i<j} \dfrac{1}{\vert \bs{r}_i-\bs{r}_j\vert}+b.g.,
\end{equation}
where the canonical momentum of particle $i$ is $\bs{\Pi}_i=\bs{p}_i-e\bs{A}(\bs{r}_i)$ with Landau gauge chosen for the vector potential $\bs{A}(x, y)=(-B y, 0)$. The final term, $b.g$, represents the neutralizing charge background. At the field strengths considered here, the Zeeman energy is large enough that the electrons are fully spin-polarized and we treat them as spinless throughout. 
The moir\'e potential depth is characterized by $\lambda=V_m/W$ and the electron-electron interaction strength is tuned by $r_s=a/a_B^*$.
We quote the depth $V_m$ in meV throughout, so that $\lambda=V_m r_s^2/\mathrm{Ha}^*$.
Here $a_B^*=a_B\,\epsilon_r/(m^*/m_e)$ is the effective Bohr radius, with $a_B$ its vacuum
counterpart, $m_e$ the free-electron mass and $\epsilon_r$ the relative permittivity of the
screening environment, and $\mathrm{Ha}^*=\mathrm{Ha}\,(m^*/m_e)/\epsilon_r^2$ is the
corresponding effective Hartree.
We use the leading-order approximation to the moir\'e potential in reciprocal space  $\Lambda(\bs{r})=\sum\limits_{j=1}^3 2\cos(\bs{r}\cdot\bs{g}_j+\phi)$, where $\bs{g}_j$ are three of the smallest non-zero reciprocal lattice vectors of the moir\'e unit cell.
The shape of the moir\'e potential is controlled by a single parameter, $\phi$,
where $\phi=60^\circ$ creates a honeycomb lattice with two degenerate local minima per moir\'e unit cell, whereas $\phi=0^\circ$ generates a triangular lattice with one unique minimum per moir\'e unit cell.

As highlighted in Fig.~\ref{fig:1}, $\phi=20^\circ$ and $\phi=45^\circ$ have been realized experimentally and represent the strongly and weakly sublattice biased geometries. The case of $\phi=20^\circ$ maps to the Wigner molecular crystals observed in experiment~\cite{li2024wigner}. 
The perfect honeycomb geometry, $\phi=60^\circ$, is also of particular interest, as we discuss below, because the three electrons per cell distribute themselves over singly and doubly occupied sites with doublons realizing short range order.

We map our model to the MoSe$_2$/WSe$_2$ heterobilayer, where the lattice constants of the two layers are approximately equal to $a_0=3.28$~\AA.
The effective mass of electrons doped into the MoSe$_2$ layer is $m^*\approx 0.5\,m_e$.
Assuming the relative permittivity of an hBN-encapsulated device to be approximately $\epsilon_r\approx 5$, we get an effective Bohr radius roughly $10$ times larger than its vacuum counterpart, $a_B^*=10\,a_B=0.529$~nm, and an effective Hartree $\mathrm{Ha}^*=544.2$~meV.
We investigate conditions within the capabilities of the experiment by targeting twist angles $\theta\approx0.54^\circ, 0.72^\circ,$ and $1.08^\circ$, 
which correspond to moir\'e periods of $a_M\approx a_0/\theta=34.9, 26.2,$ and $17.5$ nm, respectively.
With three electrons per moir\'e cell ($\nu_m=3$), the effective interaction strength as measured by the Wigner-Seitz parameter is $r_s=20, 15,$ and $10$.
Unless stated otherwise we work at $r_s=15$ ($a_M\approx26$~nm, $B\approx21$~T), and vary $r_s$ only in Sec.~\ref{sec:results} where the doublon ordering transition is located.
The magnetic field strengths needed to achieve $\nu_B=1$ are $B=\dfrac{2}{\nu_B r_s^2} \left( \dfrac{\hbar}{e}/{a_B^*}^2\right)=11.75, 20.89,$ and $47$ T.

\section{Method}
\label{sec:method}
To solve the above model accurately presents an outstanding challenge. Even the treatment of the 2DEG itself, which requires balancing the quantum zero-point motion and strong long-range interaction, already pushes the limit of state-of-the-art many-body computation~\cite{Smith_2024}. The introduction of the external moir\'e potential and especially the magnetic field provide two new elements of competition. Most common approximations, such as mean-field or perturbative approaches, or simplifications such as a lattice model with projection to one or a few bands, do not work well. The ability to describe all aspects of this problem with quantitative accuracy is very limited and would benefit greatly from the development of more accurate and scalable methods.

Variational Monte Carlo~\cite{McMillan1965, Foulkes2001} with neural quantum states (NQS) is one such method and it provides a systematically improvable variational
description of correlated continuum
electrons~\cite{Carleo2017,Pescia_2024,Smith_2024}, but they have not been
used to map phase diagrams of magnetic systems, for three reasons. First,
the torus geometry is preferred for ground-state studies due to its ability to
accommodate competing orders and extrapolate to the
thermodynamic limit. However, in the presence of a magnetic field, wavefunctions must obey magnetic periodic
boundary conditions, which standard continuum ans\"atze do not satisfy.
Second, a topological invariant is a property not of one wavefunction but
of a family of ground states spanning a cycle of boundary conditions, so
measuring it directly has often required separate computations at different twist angles, which presents additional challenges, especially in the presence of stochastic noise. Third, resolving a phase boundary requires many such calculations
across Hamiltonian parameter space. 
Here we present a framework that addresses these in turn:
a fully periodic neural-network multiplicative orbital backflow
acting on a Landau-level basis
(Sec.~\ref{sec:prefactor}) satisfies the boundary conditions by
construction, while conditioning the same network on Hamiltonian parameters
and on the boundary twist (Sec.~\ref{sec:chern}) collapses what would be many independent optimizations (nearly a thousand in the present case)
into a small number of foundation
models. These ingredients make the phase diagram and its topological
classification within reach, with calculations which are both highly accurate and affordable.

\subsection{Neural-network Multiplicative Orbital Backflow for correlated Landau-level states}
\label{sec:prefactor}
\begin{figure}[!htp]
  \centering
  \includegraphics[width=\linewidth]{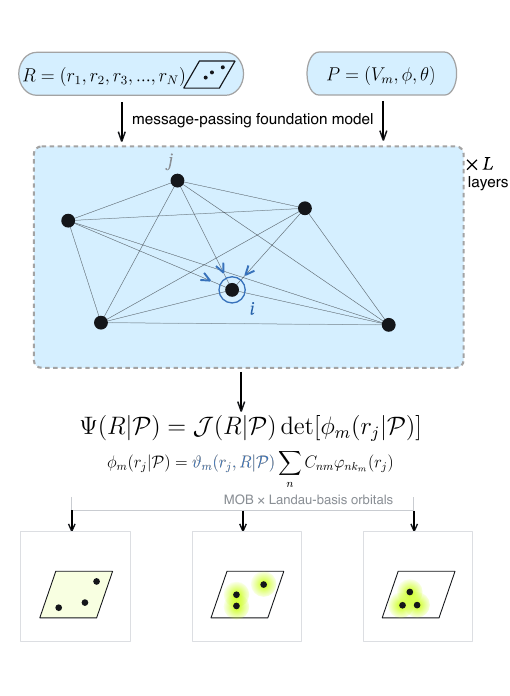}
  \caption{Summary of computational approach: ansatz and
  architecture of the foundation neural quantum state. Electron coordinates and the Hamiltonian parameters conditioning vector $\mathcal{P}$ enter a message-passing attention block, whose output is the periodic multiplicative orbital backflow (MOB) prefactor $\vartheta_m$ multiplying the Landau-level basis in Eq.~(\ref{eq:prefactor}), which satisfies MPBC.
  }
  \label{fig:NN}
\end{figure}
In Fig.~\ref{fig:NN} we illustrate the foundation neural network ansatz we use throughout this work. The neural network itself is contained within the message-passing attention block and is similar to the MPNQS~\cite{Pescia_2024, Smith_2024} architecture used to study the 2DEG and moir\'e systems~\cite{smith2026neuralnetworkdiscoverypaired}. The key difference is the conditioning on Hamiltonian parameters and that, in place of traditional backflow, we adopt a \textit{periodic neural-network, multiplicative orbital-specific backflow} (NN-MOB) in tandem with a Landau-level basis. This construction allows the network to encode correlations directly at the many-body level.
The distinction is particularly important in the present setting, where the interplay of magnetic fields, interactions, and moir\'e potentials generates highly nontrivial correlation structures and magnetic periodic boundary conditions which are naturally satisfied in this framework. This prefactor-based architecture also provides greater expressivity due to its ability to modify every orbital independently. Similar approaches to enforcing the magnetic periodic boundary condition have recently been used to study the $\nu_B=1/3$ fractional quantum Hall state~\cite{sb8g-qvn6,9qlr-jp6x}.

We work in the Landau gauge with $A(x,y) = (-By,0)$ for simplicity. In this gauge, the Hamiltonian is invariant under translations along $x$, but translations along $y$ must be accompanied by a gauge transformation. The magnetic periodic boundary condition (MPBC) on a cell of dimension $L_x \times L_y$ that is threaded by an integer number of flux quanta, $N_\phi=L_xL_y/(2\pi \ell^2_B)$ with the characteristic magnetic length, $\ell_B=\sqrt{\hbar/(eB)}$, is given by:
\begin{align}
    \psi(x+L_x,\,y) &= \psi(x,y), \nonumber\\
    \psi(x,\,y+L_y) &= e^{-iL_y x/\ell_B^2}\,\psi(x,y).
\end{align}
The noninteracting eigenstates satisfying this condition are, up to normalization, the Landau orbitals: 
\begin{align}
\varphi_{n,k_m}(x,y)\propto\;& \sum_{p\in\mathbb Z} e^{i\left(k_m - pL_y/\ell_B^2\right) x}
\exp\!\left[-\frac{(y-Y_m-pL_y)^2}{2\ell_B^2}
\right]\nonumber\\
&\times H_n\!\left(\frac{y-Y_m-pL_y}{\ell_B}\right),
\end{align}
where $n$ denotes the Landau level index, $k_m=2\pi m/L_x$ is the x-momentum associated with the guiding center $Y_m=-k_m\ell_B^2$, the sum over images $p$ enforces MPBC and $H_n$ are the Hermite polynomials. Landau levels can be combined and used as an orbital basis~\cite{Reddy_Qmelt-WC-Hall_2025} which is the approach we take. 

Our variational wave function ansatz is written explicitly  in Fig.~\ref{fig:NN}.
Orbital $m$  in our Slater determinant is 
given by:
\begin{equation} \label{eq:prefactor}
    \phi_m(r_j) =  \vartheta_m(r_j,\vec{r}\,|\mathcal{P}) \sum_{n=0}^{n_{\rm LL}-1} W_{mn}\, \varphi_{n,k_m}(r_j)
\end{equation}
where there is one orbital per guiding center $k_m$, the multiplicative orbital
backflow (MOB) $\vartheta_m(r_j,\vec{r}\,|\mathcal{P})$ is a fully periodic
neural-network function of all particle positions, conditioned on a
vector of external parameters $\mathcal{P}$, and the sum over $n$ mixes the lowest
$n_{\rm LL}$ Landau levels at that guiding center with learnable coefficients
$W_{mn}$. 

The
conditioning vector, $\mathcal{P}$, is what makes the ansatz a foundation model: rather
than representing the ground state of a single Hamiltonian, one set of
network weights defines a continuous family of variational states. For
the phase-diagram scans we condition on the Hamiltonian parameters,
$\mathcal{P}=(V_m,\phi)$, at fixed boundary twist; for flux-insertion
calculations we instead fix the Hamiltonian and condition on the twist,
$\mathcal{P}=\theta_x$ (Sec.~\ref{sec:chern}). In both cases the
conditioning variables are scaled to the unit interval.

The neural network that produces 
the MOB
$\vartheta_m(r_j,\vec{r}\,|\mathcal{P})$ utilizes an attention mechanism over distances  $r_{ij}$ and mixes in one-body features, $r_i$. This mechanism and message passing structure is largely similar to the one proposed in~\cite{Pescia_2024} and utilized in the two-dimensional electron gas~\cite{Smith_2024,ge2025visualizingimpactquencheddisorder} and moir\'e systems without magnetic field~\cite{smith2026neuralnetworkdiscoverypaired}. The main modification is to include the moir\'e parameters as input to the neural network alongside the electron coordinates and, instead of parameterizing backflow independent of orbitals, we parameterize the orbital prefactors. For more details see the Supplementary Material.
\subsection{Parameter conditioning and direct measurement of many-body
Chern numbers}
\label{sec:chern}

A second major bottleneck in NQS studies of topological phases is the
reliable computation of topological invariants. The many-body Chern
number is not a property of a single wavefunction but of a family of
ground states spanning a full cycle of twisted boundary conditions, so
its direct evaluation ordinarily requires many independent, converged
optimizations; most existing approaches therefore fall back on proxies
such as approximate Berry curvatures or symmetry indicators. Here we
evaluate the invariant directly. Threading one flux quantum through the
torus pumps an integer charge across the simulation
cell~\cite{Laughlin1981,Niu1985}. For each system Hamiltonian, with fixed $(V_m,\phi)$, the Chern number is given via this pumping by
\begin{equation}
\label{eq:chern}
    C = \int_0^{2\pi} d\theta_x\, \partial_{\theta_x} P_y(\theta_x),
\end{equation}
where the
Resta polarization~\cite{Resta_1998}
$P_y(\theta_x)=\frac{1}{2\pi}\Im\ln\big\langle
e^{i\frac{2\pi}{L_y}\sum_j y_j}\big\rangle$
is the dimensionless many-body polarization of the optimized state at
twist $\theta_x$. The winding of $P_y$ over one flux cycle equals the
pumped charge and hence the Chern number. This measure is fully
many-body and gauge invariant, and in variational Monte Carlo it
reduces to the phase of a single expectation value, sampled at
negligible cost alongside the energy (see the Supplemental Material for
the derivation and the discrete estimator).

Parameter conditioning is what renders this measurement---and the phase
diagram itself---affordable, and we use it in two complementary modes.
First, to map charge-ordering tendencies, we train
Hamiltonian-conditioned models with $\mathcal{P}=(V_m,\phi)$ at a fixed
boundary twist, which yield ground-state energies, densities, pair
correlations, structure factors, and sublattice occupations across the
entire two-dimensional parameter space from a single optimization.
Second, to assign Chern numbers, we train twist-conditioned models with
$\mathcal{P}=\theta_x$ at a fixed Hamiltonian, so that one model
represents the full flux-insertion cycle of the Hamiltonian
rather than requiring a
separate optimization at every twist. In practice we compute $P_y$ on a
discrete grid of twist angles, unwrap its phase continuously over the
cycle, and obtain the pumped charge from the net winding. This two step method allows us to obtain a cheap preliminary phase diagram and then determine phase boundaries more accurately with twist-based models at single points in the phase diagram.

\subsection{Observables and Correlation Functions}

We characterize charge order with three real-space diagnostics: the pair correlation function, the static structure factor, and the sublattice occupation. The charge density is defined as

\begin{equation} \label{eq:rho}
\rho(\bs{r}) = \frac{\Omega}{N}\Big\langle \sum_j \delta(\bs{r}-\bs{r}_j) \Big\rangle,
\end{equation}
where $\Omega$ is the area of the simulation cell and $N$ the number of electrons,
so that the perfectly uniform liquid is 1 everywhere.
The pair correlation function is defined as
\begin{align} \label{eq:vecgofr}
g(\bs{r}) =  \dfrac{1}{\Omega} \int d\bs{r}'\, g^{(2)}(\bs{r}', \bs{r}'+\bs{r}),
\end{align}
where
\begin{align}
g^{(2)}(\bs{r}_1, \bs{r}_2) &= \frac{\Big\langle
\sum_{i\neq j} \delta(\bs{r}_1-\bs{r}_i)\,  \delta(\bs{r}_2-\bs{r}_j)\Big\rangle}{\rho(\bs{r}_1)\,\rho(\bs{r}_2)}.
\end{align}
Because the two moir\'e sublattices are the natural reference for charge
order in the honeycomb geometry, we also resolve $g(\bs{r})$ by sublattice.
We partition the moir\'e cell into the Voronoi regions $\Omega_A$ and
$\Omega_B$ associated with the two potential minima, assign each electron
to a sublattice by the region containing it, and separate pairs into
same-sublattice and opposite-sublattice contributions,
\begin{equation}
g(\bs{r}) = g_{AA}(\bs{r}) + g_{AB}(\bs{r}),
\end{equation}
where $g_{AA}$ describes correlations between electrons on the same sublattice and $g_{AB}$ for those on opposite sublattices. Both are averaged
over the reference position as in Eq.~(\ref{eq:vecgofr}).

The same Voronoi cell procedure  is used to characterize the states as $(2:1)$ and $(3:0)$ states and we show the sublattice occupation development in the Supplemental Material. Similarly, we study the emergent Ising degree of freedom in the honeycomb state by counting the electrons within each Voronoi region and identifying it as singly or doubly occupied. 
For each moir\'e potential minimum $i$, we define the corresponding Ising variable \begin{equation} \label{eq:sigma}
\tau_i = \begin{cases} +1, & 
\text{ doubly occupied},\\ -1, & 
\text{ singly occupied},\end{cases}
\end{equation}
from which we obtain the conditional doublon probabilities and the doublon correlation function $\langle \tau_i \ \tau_j\rangle$ shown in Fig.~\ref{fig:phi0}.
The static structure factor is defined as,
\begin{equation} \label{eq:sofk}
S(\bs{k}) = \frac{1}{N}\left\langle \rho_{\bs{k}}\, \rho_{-\bs{k}} \right\rangle,
\qquad \rho_{\bs{k}} = \sum_{j=1}^{N} e^{-i\bs{k}\cdot\bs{r}_j},
\end{equation}
where $N$ is the number of electrons. Its largest value over nonzero wavevectors, $\max S(k)$, grows with the moir\'e-scale charge modulation and serves as a measure of the strength of correlation and charge ordering.

\section{Results}
\label{sec:results}
Our central results are summarized in Fig.~\ref{fig:1}, which shows the interplay of topological and charge orders in magnetic moir\'e systems with one magnetic flux quantum per electron ($\nu_B=1$) and three electrons per moir\'e unit cell ($\nu_m=3$). When the moir\'e potential is sufficiently weak, the 2D electron system remains in an integer quantum Hall (IQH) liquid state, as expected.
As the moir\'e potential deepens, charge ordering emerges. However, the type of charge order and its topological character depend strongly on the shape of the moir\'e potential ($\phi$) and its strength.
When the moir\'e potential shape is strongly triangular ($\phi<40^\circ$), all three electrons in the unit cell become localized in one potential minimum at sufficiently deep $V_m$ (Fig.~\ref{fig:1}d).
In contrast, if the potential shape is closer to honeycomb ($40^\circ < \phi < 60^\circ$), then only two electrons localize in the main minimum while one settles in the local minimum at the ``B sublattice'' to minimize Coulomb repulsion (Fig.~\ref{fig:1}c).
At the honeycomb limit ($\phi=60^\circ$), the charge density distributes evenly across the energetically degenerate A and B sublattices.
As we move from the triangular to the honeycomb geometry, the topological character is retained over a wider window of $V_m$ values, deeper into the charge-ordered phase.
Table \ref{tab:phases} provides a compact guide to the five regimes presented in this phase diagram.

\begin{table}[tb]
\caption{Description of the different regimes in Fig.~\ref{fig:1}. CI denotes Chern insulator.}
\label{tab:phases}
\begin{ruledtabular}
\begin{tabular}{p{0.19\linewidth}p{0.2\linewidth}p{0.5\linewidth}c}
Regime & Condition & Real-space character & $C$ \\
\colrule
IQH liquid & vanishing moir\'e & nearly uniform density, weak moir\'e modulation & 1 \\
Trivial Wigner molecules & deep moir\'e not at the honeycomb limit & $(2:1)$ or $(3:0)$ ``molecules'' pinned by the moir\'e potential & 0 \\
Ordered CI $(3:0)$ & triangular moir\'e & three-electron molecule on a single sublattice & 1 \\
Ordered CI $(2:1)$ & between triangular and honeycomb moir\'e & two electrons on one sublattice, one on the other & 1 \\
Correlated Doublon CI& honeycomb moir\'e & symmetric density; doublons alternate locally ($\phi=60^\circ$)& 1 \\
\end{tabular}
\end{ruledtabular}
\end{table}

We begin our discussion in experimentally realized systems (without magnetic fields) such as twisted bilayer WS$_2$ ($\phi=20^\circ$)~\cite{li2024wigner}, MoSe$_2$/WSe$_2$ ($\phi=26^\circ$)~\cite{Wu2018}, and WSe$_2$/WS$_2$ ($\phi=45^\circ$)~\cite{Regan2020,Li2021,Yang_moire-quantum-chemistry_2020}, 
comparing computed charge orders to what is known in the literature. We then present predictions for the topological nature of these charge orders in the presence of the magnetic field at $\nu_B=1$ as a function of moir\'e potential depth, for a variety of other lattice geometries.

\subsection{Charge Order and Topology}

For twisted bilayer WS$_2$ ($\phi=20^\circ$), the moir\'e potential is strongly triangular with one unique minimum per unit cell. Here, we observe the same ``Wigner molecular crystal'' as reported experimentally by Li \textit{et al.}~\cite{li2024wigner}, where three electrons occupy the same minimum, forming a lattice of ``trimers''.
In a deep moir\'e potential ($V_m \ge 5$~meV), this phase is exponentially localized and topologically trivial.
In a shallower potential, however, topology coexists with the trimer charge order and the system retains a Chern number of $C=1$.

For WSe$_2$/WS$_2$ ($\phi=45^\circ$), the moir\'e potential evolves from triangular towards honeycomb symmetry. A second set of local minima develop at what would eventually become the honeycomb B sublattice. At this shape, the B minima are shallower than those at A with a potential barrier in between.
In a deep moir\'e potential ($V_m \ge 10$~meV), the system forms a charge-transfer insulator with $2:1$ charge ratio on the A and B sublattices.
The charge on each site obeys the same $C_3$ symmetry as the potential but now the A sublattice contains only two out of the three electrons with the third localizing on the B sublattice.
Similarly to twisted bilayer WS$_2$ ($\phi=20^\circ$), the system retains topology ($C=1$)  in a shallow potential. Here, however, the topology survives to much deeper moir\'e potentials, up to $V_m\sim 10$~meV rather than $5$~meV as in the previous case.
This trend of topology surviving in deeper potentials continues as we tune the shape of the moir\'e potential further towards the honeycomb limit.
At $\phi=60^\circ$, the moir\'e potential has honeycomb symmetry with two degenerate minima per unit cell.
Remarkably, here the Chern number remains $1$ 
throughout and no topologically trivial state is seen in the range of moir\'e potential we explored.

\begin{figure}[htp!]
  \centering
  \includegraphics[width=\linewidth]{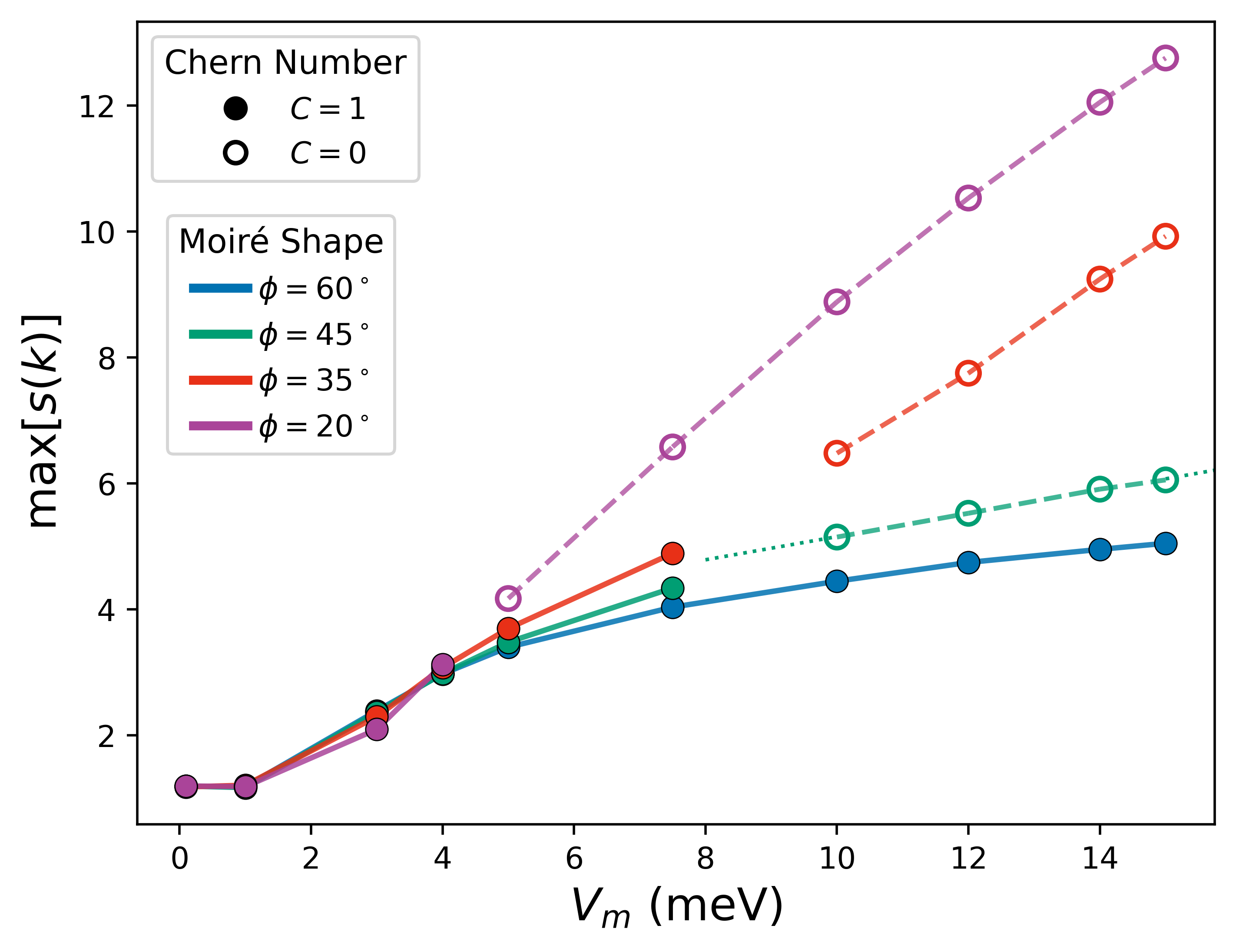}
  \caption{Quantifying charge order and topology as a function of moir\'e potential depth in select lattice geometries. The peak of the static structure factor, $\max S(k)$, is a measure of the degree of charge order. Topology is characterized by the many-body Chern number $C$.
  At each moir\'e potential shape ($\phi$), the topological sector is connected by solid lines, the trivial sector is connected by dashed lines. For $\phi=45^\circ$ we indicate, as an example, the slope with the dotted line.
 }
  \label{fig:chern_sofk}
\end{figure}

This interplay between topology and charge order is quantified in Fig.~\ref{fig:chern_sofk}, where we show the magnitude of the peak of the static structure factor, $\max S(k)$, which measures the degree of long-range charge order, along with the Chern number, $C$, which characterizes topology.
Across most moir\'e geometries, increasing $V_m$ enhances localization, as reflected in the growth of $\max S(k)$. However, localization alone does not determine the topology. There are broad regions where the density is strongly modulated but the many-body Chern number remains $C=1$.
Away from the honeycomb limit ($\phi=60^\circ$), topology disappears in a sufficiently deep moir\'e potential.
In this topologically trivial regime, the structure factor peak grows linearly with the depth of the moir\'e potential. When the Chern number is nonzero, however, the growth is suppressed. The slope of $\max S(k)$ versus $V_m$ therefore changes as the system passes between the two regimes.

\subsection{
Doublon correlations and ordering  in honeycomb moir\'e}

\begin{figure*}[!htp]
  \centering
  \includegraphics[width=\linewidth]{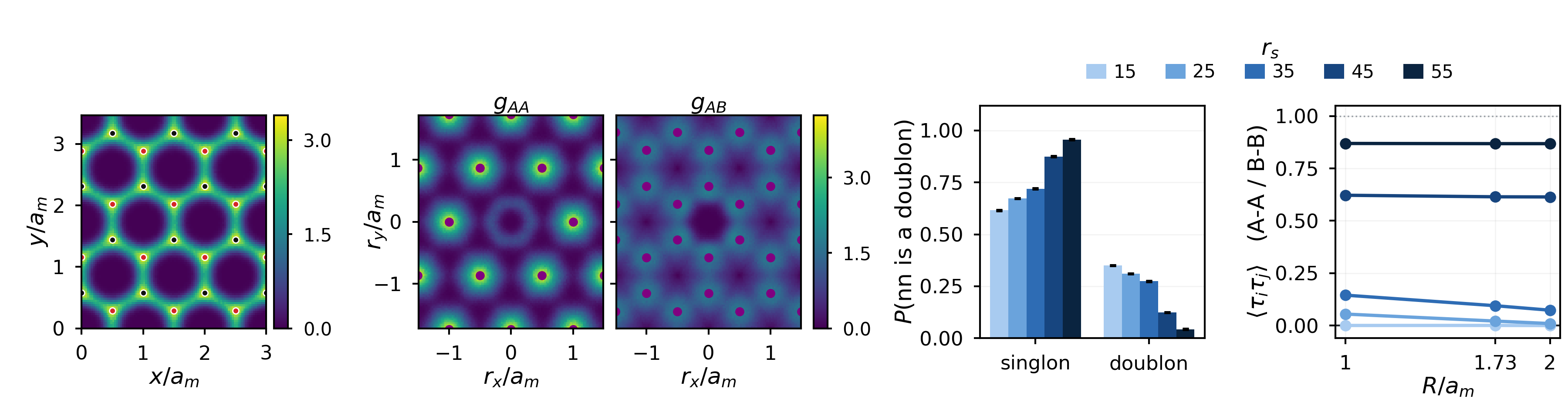}
 \caption{Doublon correlations and their ordering at $\nu_B=1$ and $\nu_m=3$ filling of the honeycomb moir\'e lattice ($\phi=60^\circ$) with a potential depth of $V_m=10$~meV. (a) Charge density at $r_s=15$, with red and black points marking the A and B sublattices.  No sublattice symmetry breaking is seen.
 (b) Sublattice resolved pair correlation functions at $r_s=15$ $g_{AA}(\bs{r})$ and $g_{AB}(\bs{r})$ with the reference electron at the origin; the purple markers correspond to the same (in $g_{AA}$) and opposite (in $g_{AB}$) sublattice minima.  The correlation shows clear sublattice ordering.
 (c) Probability that a site hosts a doublon, conditioned on its nearest neighbor being singly occupied (left) or doubly occupied (right), for interaction strengths $15\leq r_s\leq55$. (d) Same sublattice doublon correlation function $\langle \tau_i \tau_j\rangle$ versus separation $R$ in units of the moir\'e period $a_M$, for the same range of interaction strengths as in (c).
  }
  \label{fig:phi0}
\end{figure*}

In the honeycomb potential ($\phi=60^\circ$), the two minima in the moir\'e unit cell are degenerate so any $(2:1)$ arrangement of the electrons is selected by 
 correlation rather than being imposed by the moir\'e potential. We first characterize the state at conditions where we find local but not long-ranged order. We then increase the interaction strength, to establish that this local order is the precursor of a genuine long-range order. The one-body charge density alone is insufficient to characterize 
this correlated electronic state.
Figure~\ref{fig:phi0}(a) shows the charge density, which respects the sublattice symmetry, with an average of $1.5$ electrons on each sublattice per moir\'e unit cell.
Based on this and the computed Chern number, this state is essentially indistinguishable from the  IQH liquid
at very weak potential depth, 
and becomes progressively less uniform as   $V_m$ increases.
However, this symmetric density does not imply that the electrons occupy the two sublattices independently.
Instead, the occupations of neighboring sublattice sites are correlated by interaction.
Because the Coulomb energy heavily penalizes triple occupancy, the electrons distribute in a $(2:1)$ pattern over the two minima in each unit cell. Since the electrons are polarized, the doubly occupied site must host electrons in distinct orbitals in a correlated two-electron doublon. The state is then a mixture of single electrons and doublons. The natural question is whether the doublons order to maximize their respective distances.

The two-body density--density correlations in Fig.~\ref{fig:phi0}(b) already show that the doublons are correlated. The sublattice resolved pair correlation function $g(\bs{r})$, defined in Eq.~(\ref{eq:vecgofr}), gives the probability of finding a second electron displaced by $\bs{r}$ from a reference electron, averaged over the reference position. Its maxima fall on sites belonging to the same sublattice as the reference electron, while the opposite sublattice is suppressed, indicating that electrons are more likely to be found near a site of the same occupancy class than near its neighbors.

Figure~\ref{fig:phi0}(c) isolates this tendency directly, through the occupancy of a site conditioned on that of its nearest neighbor, which is on the opposite sublattice. At $r_s=15$, conditioning on a singly occupied neighbor raises the probability that a site hosts a doublon to 0.62, well above the uncorrelated value, while conditioning on a doublon neighbor lowers it to 0.35. Doublons therefore repel one another and preferentially sit adjacent to singly occupied sites. In the Ising language of Eq.~(\ref{eq:sigma}), the effective nearest-neighbor coupling is antiferromagnetic. The pattern is similar to the $(2:1)$ arrangement realized at $\phi$ slightly away from honeycomb where all doublons occupy a single sublattice. At the honeycomb limit, the choice of sublattice is degenerate and selecting one would spontaneously break symmetry.

The system does not break symmetry at this interaction strength. Figure~\ref{fig:phi0}(d) tracks the same sublattice correlation $\langle \tau_i\tau_j\rangle$ out to $R=2a_M$. At $r_s=15$, the correlation is negligible at every separation and the local alternation does not propagate beyond the nearest neighbor. Thus, at $r_s=15$, the honeycomb point realizes a Chern insulator with short-range doublon correlations.

Because the doublon-doublon repulsion is Coulombic in origin, it is natural to ask if this local order is the precursor to a phase where doublons order. We therefore investigate how doublon correlations change as we increase the interaction strength. The conditional probabilities of Fig.~\ref{fig:phi0}(c) strengthen as we increase $r_s$ with the singlon-conditioned value rising to $0.96$ and the doublon-conditioned value falling to $0.04$ by $r_s=55$. Fig.~\ref{fig:phi0}(d) shows that this correlation does not decay for $r_s\geq 45$, suggesting that the correlation has become long-range. The honeycomb point therefore supports two regimes separated by an interaction-driven ordering transition at a critical $35<r^*_s\leq45$. Below $r^*_s$, the state is a Chern insulator with short-ranged doublon correlations that grow in strength and range as we increase $r_s$.
This correlation eventually becomes long-range and spontaneously breaks sublattice symmetry, changing the state to a N\'eel CDW ~\cite{Yang_FerromagneticSemimetalChargeDensity_2024,Capponi_PhaseDiagramInteracting_2015} of doublons.
Establishing this transition more firmly will require further studies in larger system sizes.

\subsection{Topological to Trivial Transitions}

\begin{figure*}[!htp]
\centering
    \includegraphics[width=\linewidth]{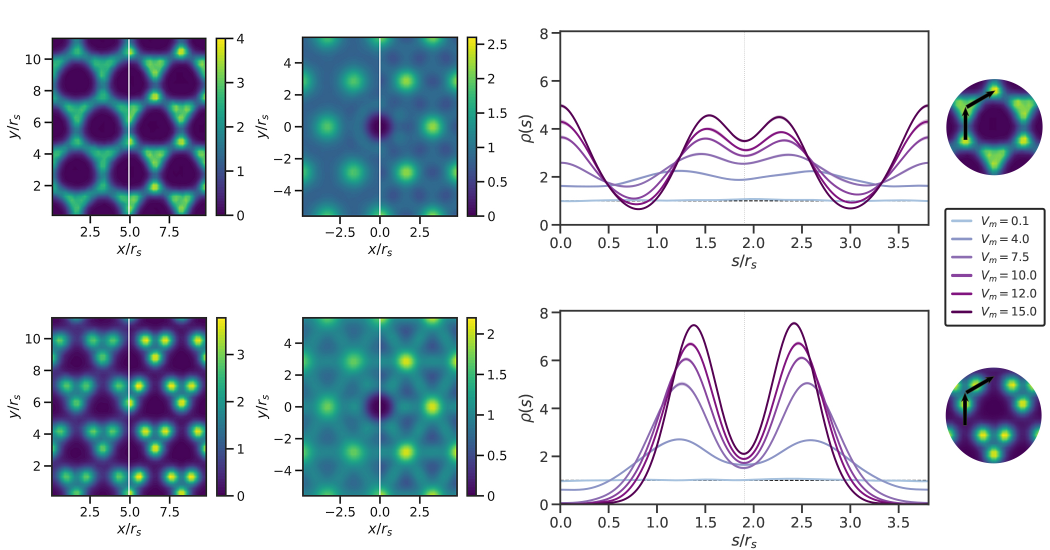}
  \caption{Charge density and pair correlation across representative transitions from $C=1$ to $C=0$.
  The top row shows the model for WSe$_2$/WS$_2$ ($\phi=45^\circ$). The bottom row shows the bilayer WS$_2$ ($\phi=20^\circ$) ground states. Each density panel is divided in two, with the left half showing the Chern number $C=1$ state and the right half showing the $C=0$ state. The third panel in each row shows linecuts of the density plots for six values of $V_m$ in meV (values shown in the legend box on the right), through the high-density locations of all three electrons in the moir\'e unit cell, as indicated by the motifs on the right. These clearly distinguish the charge-ordered states at $\phi=45^\circ$ and $\phi=20^\circ$: the former has substantial occupation on the honeycomb B-sublattice, whereas the latter has almost none.
  Note that in each sub-panel, the line for $V_m=0.1$~meV (IQH phase) is essentially indistinguishable from the 
  horizontal dotted line, which represents constant density.
  }\label{fig:combinedfig4}
\end{figure*}

We now examine how topology is lost away from the honeycomb limit when $V_m$ becomes sufficiently deep.
As $\phi$ deviates from the honeycomb value of $60^\circ$, the A and B minima become inequivalent.
The two material-relevant regimes, representative of WSe$_2$/WS$_2$ ($\phi=45^\circ$) and bilayer WS$_2$ ($\phi=20^\circ$), lie on opposite sides of the geometry-dependent crossover between two-sublattice and single-sublattice charge order.
The former retains substantial occupation of both sublattices while the latter collapses to a single triangular sublattice.
Figure \ref{fig:combinedfig4} compares representative topological ($C=1$) and trivial ($C=0$) states for these two geometries, with the left and right halves of each density and pair-correlation panel showing the two topological sectors, respectively.

At $\phi=60^\circ$, the average occupations of the A and B minima remain equal at all potential depths.
Away from this limit, the charge distribution becomes increasingly sublattice polarized as the moir\'e potential deepens.
Depending on the potential geometry, the occupations approach either a $(2:1)$ or a $(3:0)$ pattern.
At $\phi=20^\circ$, all three electrons occupy the same A-sublattice minimum, forming a $(3:0)$ trimer. The density linecut shows that the trimer remains $C_3$-symmetric within numerical noise.
Each trimer is locked to the same orientation as dictated by the moir\'e potential.
The principal effect of increasing the potential depth is therefore to sharpen an already established molecular charge pattern.
At $\phi=45^\circ$, the system instead forms a $(2:1)$ configuration, with approximately two electrons associated with the deeper A-sublattice minimum and the third associated with the B-sublattice minimum.
Notably, the density surrounding the doubly occupied A minimum retains approximate $C_3$-symmetry, rather than developing a $C_2$-symmetric, two-lobed structure.
This indicates that the two electrons remain delocalized within the moir\'e minimum rather than occupying two fixed locations, as predicted by the Hartree–Fock picture~\cite{li2024wigner}.
Correlations allow this delocalization while avoiding a large increase in the Coulomb energy.

The pair-correlation functions provide a complementary view of this evolution.
Their brightest peaks, arranged on a triangular lattice, arise primarily from same-sublattice correlations, while the weaker features at shorter separations contain contributions from opposite-sublattice pairs.
At $\phi=20^\circ$, the overall correlation pattern changes little across the topological transition and mainly becomes sharper as the moir\'e potential deepens. Thus, the loss of topology occurs without a qualitative reorganization of the local three-electron molecule.
The evolution at $\phi=45^\circ$ is more pronounced.
In a shallow potential, where $C=1$, the electrons remain significantly delocalized, and the weak opposite-sublattice contribution in $g(\bs{r})$ is nearly rotationally isotropic.
In the deeper, topologically trivial state, the same contribution develops a clearer sixfold pattern, signaling stronger locking of the relative electron positions to the moir\'e lattice. Nevertheless, the density and pair correlations evolve continuously rather than displaying an abrupt local signature at the Chern transition.

These comparisons establish that the loss of topology need not coincide with a qualitative change in charge order. Both the amplitude of the density modulation and the degree of sublattice polarization vary smoothly as the many-body Chern number changes from $C=1$ to $C=0$.
Local observables can therefore look similar on opposite sides of the transition, underscoring the need to evaluate the many-body topological invariant directly.

\section{Discussion}
\label{sec:discussion}
Our results establish how  the Wigner-molecular physics already accessible in TMD moir\'e experiments manifests and generalizes in the presence of a magnetic field. At weak potential, the three electrons per cell form an integer quantum Hall liquid. At strong, triangular confinement they approach the topologically trivial three-electron molecule imaged at zero field~\cite{li2024wigner}. Between these limits lies a broad and experimentally accessible regime in which the molecular charge distribution is already pronounced but the Chern number remains $C=1$.

Our results reveal that geometry and topology are intertwined, but localization and topology are not equivalent. Similar $(2:1)$ or $(3:0)$ density patterns occur on both sides of the Chern transition, and a state can be partially polarized into one potential minimum while remaining topological. 
Conversely, topology can be lost with only a small change in local observables. The systematic movement of the phase boundary with $\phi$ shows that the potential landscape controls the stability of the Chern state, while the absence of a universal local signature shows why a direct many-body invariant is indispensable.

This geometry dependence has a natural interpretation in terms of
strong-coupling limits. At large moir\'e depth, a biased potential is adiabatically
connected to a product of well-separated three-electron molecules---a
trivial insulator---so a transition out of the $C=1$ state must occur at
some finite depth, and it occurs sooner the more directly the geometry
approaches that limit. The honeycomb potential admits no such limit: three
electrons cannot be divided symmetrically between two equivalent minima, so
no symmetric, site-localized insulator exists to which the ground state
could connect. The system instead exhibits a local correlation where doublons prefer to alternate with single electron sites.
This is similar to the long-range order seen at $\phi=50^\circ$, but it remains short-ranged and the
 Chern number remains quantized.
As the potential depth grows, 
the range of the order increases,
until sublattice symmetry is broken and long-range N\'eel CDW order
develops.

These predictions are testable with existing capabilities. For the largest
moir\'e period we study ($a_M\approx35$~nm, $\theta\approx0.54^\circ$), the
$\nu_B=1$ condition corresponds to $B\approx12$~T, well within DC-magnet
range; the smallest period requires $47$~T. The topological molecular
regime would manifest as a quantized $\sigma_{xy}$ plateau coexisting with
the molecular charge pattern accessible to scanning-probe
imaging~\cite{li2024wigner,Xiang2025}. The honeycomb state requires a complementary
strategy: because the one-body density is symmetric and featureless with respect to the doublons, only density--density correlations would
reveal the local order directly. More accessible is the ordering transition in the vicinity of the doublon liquid. A weak sublattice bias from strain, displacement
field, or a small departure from the honeycomb condition would select one sublattice and convert the liquid into the $(2:1)$ pattern which is directly visible in the one-body density. One could also increase interaction strength by decreasing screening or increasing the moir\'e period. The
quantized Hall response also survives deep into the charge-ordered
regime and cleanly separates these states from trivial Wigner molecular
crystals in transport.

Several caveats and extensions merit comment. The many-body Chern number we compute characterizes the ground state on the torus, and its identification with the measured Hall conductance requires charge ordered states to be pinned: for a crystal free to slide, the center-of-mass motion contributes an additional, generally non-quantized term to $\sigma_{xy}$~\cite{ZengMillis2025}. Here the charge pattern is locked to the moir\'e potential rather than translating spontaneously, so the sliding mode is absent and the quantized internal contribution is the one probed in transport. The doublon correlations reported
here extend to separations of $2\,a_M$, which establishes that the alternation is
short ranged on that scale but cannot exclude a small long-range component; a
finite-size scaling of the doublon structure factor at the ordering wavevector
would settle whether the liquid survives in the thermodynamic limit, and combined
with the $r_s$ dependence of Fig.~\ref{fig:phi0}(d) would locate the transition
into the $(2:1)$ crystal.

Beyond $\nu_B=1$, the framework developed here applies directly to fractional
Landau-level fillings, where the interplay between fractionalized topological
order and molecular charge order remains open, and to the melting of these states
with temperature and disorder---processes whose zero-field counterparts are
already being imaged in the same materials~\cite{Xiang2025,ge2025visualizingimpactquencheddisorder}.
Finally, the methodological ingredients of Sec.~\ref{sec:method} are not specific to this
problem. Any Hamiltonian parameter can take the place of $V_m$ and $\phi$ in the
conditioning, so that one trained model resolves a continuous phase diagram rather than a
set of isolated points, and twist conditioning turns that same model into a probe
of the topological response.

\section{Conclusion}
\label{sec:conclusion}
We have established the existence of an extended topological molecular regime in which long-range molecular charge order and a quantized Hall response, $C=1$, coexist over a wide range of moir\'e depth and geometry. More generally, we have mapped the magnetic-field phase diagram of three-electron moir\'e artificial atoms at $\nu_B=1$ and $\nu_m=3$. The phase diagram connects an integer quantum Hall liquid to topologically trivial molecular states, sometimes called Wigner molecular crystals, through an extended regime of topological molecular charge ordered states. Its location at realistic moir\'e periods and magnetic fields provides a direct route from existing zero-field imaging experiments to a combined transport and real-space probe of charge order with $C=1$.

We have shown that the moir\'e geometry controls the stability of topology even when the density changes little. In the honeycomb geometry, the $(2:1)$ arrangement is chosen by the electrons rather than imposed by the potential, since the minima are degenerate. For a wide range of interactions, doublons and singly occupied sites alternate locally without establishing long-range order. 
At sufficiently strong interactions ($r_s\ge \sim 45$), 
this
Chern insulator evolves into a N\'eel charge density wave, where the $(2:1)$ pattern is established and exhibits long-range order. More generally, the results establish geometry as an independent control knob for selecting among quantum Hall liquids, topological Wigner-molecular states, and trivial electron solids.

On the methodological side, we have introduced a set of advances that allow neural quantum states to become a practical tool for discovering and characterizing topological phases. We showed how to efficiently map out a multidimensional phase diagram and resolve its topological structure by
combining (i) the integration of neural-network multiplicative orbital backflow into the Landau basis, (ii) direct evaluation of the many-body Chern number via Resta polarization, and (iii) a foundation model trained across Hamiltonian parameters and boundary conditions. These developments overcome key limitations that have previously restricted NQS to isolated benchmarks. More generally, the approach advances the state-of-the-art of 
many-body computation and establishes a scalable framework for exploring correlated quantum matter in the presence of magnetic fields.

\section{Acknowledgments}
The Flatiron Institute is a division of the Simons Foundation. The authors would like to thank Tomohiro Soejima for insightful discussions.
YY acknowledges support by the National Science Foundation (NSF) under grant number DMR-2532734. This research used resources of the National Energy Research Scientific Computing Center (NERSC), a Department of Energy Office of Science User Facility under Contract No. DE-AC02-05CH11231 using NERSC award DDR-ERCAP0038723 and the Hofstra Star HPC cluster, which was funded by the NSF Major Research Instrumentation (MRI) program (grant No. 2320735).

\bibliography{ref}
\end{document}